\documentclass{aa}  

\usepackage{graphicx}
\usepackage{placeins}
\usepackage{txfonts}
\usepackage{color}
\usepackage[table, svgnames, dvipsnames]{xcolor}
\usepackage{hyperref}
\hypersetup{colorlinks=true, citecolor=blue, linkcolor=blue}
\begin{document}

   \title{Chemical Evolution of R-process Elements in Stars (CERES)}

   \subtitle{IV. An observational run-up of the third r-process peak with Hf, Os, Ir, and Pt }

   \author{Arthur Alencastro Puls
          \inst{1},
          Jan Kuske
          \inst{2}, 
          Camilla Juul Hansen
          \inst{1}, 
          Linda Lombardo
          \inst{1}, 
          Giorgio Visentin
          \inst{3,4}, 
          Almudena Arcones
          \inst{2,4,5}, 
          Raphaela Fernandes de Melo
          \inst{1}, 
          Moritz Reichert
          \inst{6},
          Piercarlo Bonifacio
          \inst{7},
          Elisabetta Caffau\inst{7}
          \and
          Stephan Fritzsche\inst{3,4,8}
          }

   \institute{Institute for Applied Physics, Goethe University Frankfurt, Max-von-Laue-Str. 12, Frankfurt am Main 60438, Germany\\
              \email{AlencastroPuls@iap.uni-frankfurt.de}
         \and
             Institut f\"{u}r Kernphysik, Technische Universit\"{a}t Darmstadt, 64289 Darmstadt, Germany
         \and
             Helmholtz-Institut Jena, Fr\"obelstieg 3, 07743 Jena, Germany
         \and
             GSI Helmholtzzentrum f\"{u}r Schwerionenforschung GmbH, 64291 Darmstadt, Germany
        \and
             Max-Planck-Institut für Kernphysik, Saupfercheckweg 1, Heidelberg 69117, Germany
         \and
             Departament d’Astronomia i Astrof\'{i}sica, Universitat de Val\`{e}ncia, Edifici d’Investigaci\'{o} Jeroni Munyoz, C/ Dr. Moliner, 50, 46100 Burjassot, Val\`{e}ncia, Spain
         \and
             GEPI, Observatoire de Paris, Universit\'{e} PSL, CNRS, 5 place Jules Janssen, 92195 Meudon, France
         \and
             Institute for Theoretical Physics, Friedrich-Schiller-University Jena, 07743 Jena, Germany
             }

   \date{Received October 09, 2024; accepted November 18, 2024}

 
  \abstract
   {The third r-process peak (Os, Ir, Pt) is poorly understood due to observational challenges, with spectral lines located in the blue or near-ultraviolet region of stellar spectra. These challenges need to be overcome for a better understanding of the r-process in a broader context.}
   {To understand how the abundances of the third r-process peak are synthesised and evolve in the Universe, a homogeneous chemical analysis of metal-poor stars using high quality data observed in the blue region of the electromagnetic spectrum ($<$~400~nm) is necessary. We provide a homogeneous set of abundances for the third r-process peak (Os, Ir, Pt) and Hf, increasing by up to one order of magnitude their availability in the literature. }
   {A classical 1D, local thermodynamic equilibrium (LTE) analysis of four elements (Hf, Os, Ir, Pt) is performed, using \texttt{ATLAS} model atmospheres to fit synthetic spectra in high resolution ($>$ 40,000), high signal-to-noise ratio, of 52 red giants observed with UVES/VLT. Due to the heavy line blending involved, a careful determination of upper limits and uncertainties is done. The observational results are compared with state-of-the-art nucleosynthesis models.}
   {Our sample displays larger abundances of Ir (Z$=$77) in comparison to Os (Z$=$76), which have been measured in a few stars in the past. The results also suggest decoupling between abundances of third r-process peak elements with respect to Eu (rare earth element) in Eu-poor stars. This seems to contradict a co-production scenario of Eu and the third r-process peak elements Os, Ir, and Pt in the progenitors of these objects. Our results are challenging to explain from the nucleosynthetic point of view: the observationally derived abundances indicate the need for an additional early, primary formation channel (or a non-robust r-process).}
   {}

   \keywords{Stars: abundances --
                Nuclear reactions, nucleosynthesis, abundances --
                Stars: Population II
               }

   \titlerunning{CERES IV -- Hf, Os, Ir, and Pt}
   \authorrunning{Arthur Alencastro Puls et al.}
   \maketitle
%

\section{Introduction}\label{sec:introduction}

To date we do not know if the rapid neutron-capture process (r-process) always produces heavy elements in a universal, robust way. The r-process is responsible for forming half of the heavy elements, hence its nature is important to accurately understand various enrichment channels, astrophysical formation sites, as well as Galactic chemical evolution of our and other galaxies.

Since \citet[][]{Burbidge:1957aa} and \citet[][]{1977ApJ...213..225L}, both observations and theory have made immense progress, yet open questions on the exact formation site and the physics of the r-process remain. The r-process requires environments with very high neutron densities. In the past decades, the proposed locations hosting the r-process have converged to events such as rare magneto-rotational-driven core-collapse supernovae (MRSN), neutron star-black hole mergers, and binary neutron star mergers (NSM) \citep[][]{Metzger:2010aa,Winteler:2012aa,Barnes_2021,Perego_2021,Arcones_2022,Reichert_2024}.

So far, direct observations have corroborated only the NSM scenario, after detection of the r-process in the kilonova AT\,2017gfo \citep[][]{Soares_Santos_2017,Watson_2019}. That leads to the question: how unique and robust is the r-process? A NSM-only scenario is challenged by results from indirect evidence, i.e., the chemical profile of stellar photospheres, which act as the fossil record of the cloud from which they formed \citep[][]{Freeman:2002aa}. When very low metallicities are reached ([Fe/H]\footnote{[A/B] = $\log \frac{N_{A}\:N_{B,\sun}}{N_{B}\:N_{A,\sun}}$, where $N_i$ is the column density of species $i$.} $\lesssim$ $-$2.5), it has been observed that the [Eu/Fe] ratio displays a large scatter \citep[][]{Sneden:2008aa}. That result in particular is one challenge to a single site scenario, because Eu is thought to be a near-pure r-process element, at least in the Solar System \citep[][]{Simmerer:2004aa}.

Also, when the detailed composition of metal-poor stars is unveiled, it is possible to separate the stars into two groups, one lanthanide-poor and the other lanthanide-rich - even when both patterns display similar abundances of species of the first r-process peak \citep[][]{Hansen:2014aa}. Such variability contradicts the proposed robustness of the r-process. It could be explained either by an astrophysical site with a large variability of conditions or a mixture of at least two different sites. Also, the metal-rich tail of the Milky Way disc show clear indications for the need of two different formation sites, or at least sites with different delay times, to accurately explain the Eu abundances in the Milky Way disc, which also seems to be the case in dwarf galaxies \citep{Cote_2019,Molero_2021}.

In order to look for clues for solving that riddle, it is useful to look at the poorly-studied third r-process peak elements (Os, Ir, Pt), and compare their behaviour against Eu or other r-process elements between the second and third peak in low metallicity stars. These species lack a volume of data in the literature \citep[see, e.g., fig.~32 from][]{Kobayashi:2020} due to observational challenges, as their detection rely on efficient blue spectrographs with a high resolution and a high signal-to-noise ratio. Their strongest spectral lines are found in the near-ultraviolet, and often require space-based telescopes to be observed. The spectral coverage offered by the Hubble Space Telescope has been a major contributor for observational work on the r-process \citep[e.g.,][]{Sneden:2003aa,Cowan:2005aa,Roederer_2010,Barbuy_2011,Roederer_2014b,Roederer_2022}.

The combination of high-resolution, high signal-to-noise ratio (S/N) and low metallicity of the the Ultraviolet and Visual Echelle Spectrograph (UVES) spectra targeted within the Chemical Evolution of R-process Elements in Stars (CERES) survey allows the selection and analysis of a few Os, Ir, and Pt lines. In addition, we have also chosen to analyse Hf, the first post-lanthanide element between the second and third peak (atomic number, Z=72), to compare to the behaviour of the heavy r-process elements outside the third peak. So far, to our knowledge the only large survey including third r-process peak elements is the work from \citet[][]{Roederer:2014ac}, which presents 9 abundances and 61 upper limits for \ion{Ir}{I}. Here, we aim to expand the number of abundances derived homogeneously for Ir, and also to present homogeneous abundances for the other elements of the third r-process peak Os and Pt. This study is the fourth in the CERES survey, which aims to produce a full chemical profile of a sample of 52 metal-poor stars, focusing on the r-process. In \citet[][ hereafter Paper~I]{Lombardo_2022}, the sample was first presented, as well as their stellar parameters and abundances from Na to Zr. In \citet[][ hereafter Paper~II]{Fernandes_de_Melo_2024}, CNO and Li abundances are studied, while in \citet[][ hereafter Paper~III]{Lombardo_2024}, species from Ba to Eu are measured in the CERES sample.

In Sect.~\ref{sec:datanalysis} we summarise the CERES sample and the work done in Paper~I to derive the model atmospheres. We also explain in detail the procedures to fit the synthetic spectra and to separate detections from upper limits, and we also discuss the careful error analysis done in this work. Results are presented in Sect.~\ref{sec:results}, followed by a discussion both from the observational point of view, as well as their implications for the current state-of-the-art nucleosynthesis models. We make our final remarks in Sect.~\ref{sec:conclusions}.

\section{Data and analysis}\label{sec:datanalysis}

Our sample consists of 52 metal-poor red giants, with metallicities in the -3.5 $<$ [Fe/H] $<$ -1.7 interval, and it is fully described in Paper~I. None of the targets are found to be C-enhanced (Paper~II), and their r-process content is very heterogeneous, with A(Eu)\footnote{$A(X) =\,\log \frac{N_X}{N_H} + 12$. That quantity is also referred in the literature as $\log \epsilon$~(X).} abundances spanning $\approx$~2~dex (Paper~III).

The targets were observed with UVES, mounted at the Very Large Telescope \citep[][]{Dekker:2000aa}. Observations were carried out in November 2019 and March 2020 (ProjectID: 0104.D$-$0059(A), PI: Hansen). As stated in Paper~I, resolution is $R = \lambda/\Delta\lambda \gtrsim$~40,000, resulting from a configuration using an 1\arcsec~slit, 1 $\times$ 1 binning, and the standard Dic 1 configuration with blue and red arms centred on 390 and 564~nm. In this configuration, the blue CCD covers a spectral range between 326 and 454~nm.

In addition to the observations, spectra of similar quality, observed with UVES, available in the European Southern Observatory (ESO) Archive complemented the sample. The archival spectra spans resolutions ranging between 40,000 and 80,000, and a few of these targets have the blue arm centred at 346~nm. For these spectra, the wavelength coverage in the blue CCD spans from 303 to 388~nm. The resulting median signal-to-noise ratio at 390~nm is $\approx$~100 for the whole sample. Further details on the observations are described in Paper~I.

\subsection{Model atmospheres and chemical abundances}\label{sec:chem}

In order to derive abundances of Hf, Os, Ir, and Pt, we used the November 2019 version of \texttt{MOOG}\footnote{\url{https://www.as.utexas.edu/~chris/moog.html}} \citep[][]{Sneden:1973phd} to fit synthetic spectra generated with 1D, local thermodynamic equilibrium, plane-parallel, alpha-enhanced atmospheric models created with the \texttt{ATLAS12} code \citep[][]{Kurucz_2005}. These models were originally calculated for Paper~I, and they were also employed in Papers II and III to maintain homogeneity.

Stellar parameters used to create atmospheric models are those derived in Paper~I, where the iterative procedure adopted is fully described. In summary, effective temperatures T$_{\mathrm{eff}}$ and surface gravity values in terms of log~g were calculated following the method outlined in \citet[][]{Koch_Hansen_2021}, which interpolates Gaia ($G_{BP}$-$G_{RP}$) colours in a grid of model atmospheres, using the reddening law from \citet[][]{Fitzpatrick_2019}, the reddening maps from \citet[][]{Schlafly_2011}, and Gaia EDR3 photometry in the $G$, $G_{BP}$, and $G_{RP}$ bands \citep[][]{GaiaDR1:2016,GaiaEDR3:2021}. Microturbulent velocities $v_t$ were calculated using the formula from \citet[][]{Mashonkina_2017}, and the metallicities are the \ion{Fe}{I} abundances giving [Fe/H]. We refer to Paper~I for \ion{Fe}{I} and \ion{Fe}{II} abundances. The adopted uncertainty in [Fe/H] is 0.13~dex, following Paper~I.

For a given star with a given atmospheric model, the abundances A(X) were determined applying chi-square minimization between the observed and synthetic spectra for each star. To differentiate abundance detections from upper limits we developed an empirical method described in the next section.

\subsubsection*{A definition for upper limits}\label{sec:upper_limits}

Since several lines measured in this work are heavily blended, a careful treatment is necessary for flagging upper limits. After deriving the best-fitting abundance, the model of a single, isolated spectral line with the atomic parameters of the line being measured (wavelength, lower excitation potential, log~gf), as well as that best-fitting abundance, was synthesised with \texttt{MOOG}. To flag a measurement as a detection instead of an upper limit, we consider the line depth of a single line (that is synthesised alone, isolated from its blends) in a normalised spectrum with its best-fitting abundance to be expected to surpass a 3-$\sigma$ threshold over continuum noise.

To define that threshold we assume the spectral line to have a Gaussian profile. As such, for a fixed full width at half maximum, the relationship between line depth and the area of the Gaussian (i.e., the theoretical equivalent width) is linear. Hence, the treatment may be done in terms of the equivalent width $W_{\lambda}$. The formula from \citet[][]{Cayrel:1988} is adopted here to estimate the uncertainty in $W_{\lambda}$:

\begin{equation}\label{eq:cayrel}
    \sigma_{W_{\lambda}} = \frac{1.5}{S/N}\sqrt{\Delta\lambda\;\delta\lambda},
\end{equation}

\noindent where $\Delta\lambda$ is the full width at half maximum of the spectral line, $\delta\lambda$ is the pixel size, and the signal-to-noise ratio per pixel $S/N$ was estimated as the inverse of the root-mean-square of the continuum noise near each spectral line evaluated.

From Eq.~\ref{eq:cayrel} we can define a value $W_{\lambda,min}$, which corresponds to the 3-$\sigma$ threshold, as being three times the uncertainty in $W_{\lambda}$:

\begin{equation}\label{eq:threshold0}
    W_{\lambda,min} \equiv \frac{4.5}{S/N}\sqrt{\Delta\lambda\;\delta\lambda}.
\end{equation}

Dividing Eq.~\ref{eq:threshold0} by the central wavelength of the spectral line $\lambda$ and taking the logarithm on both sides, we can write it in terms of the reduced width $RW$:

\begin{equation}\label{eq:threshold1}
    RW_{\mathrm{min}} \equiv \log \frac{W_{\lambda,min}}{\lambda} = \log \left [ \frac{4.5}{\lambda\;(S/N)}\sqrt{\Delta\lambda\;\delta\lambda} \right ].
\end{equation}

The relationship shown above works for unblended lines. If we add a term $\phi$ to take into account the influence of line blending, the final equation becomes:

\begin{equation}\label{eq:threshold2a}
    RW_{\mathrm{min}} = \log 4.5 - \log \lambda - \log S/N + 0.5(\log \delta\lambda + \log \Delta\lambda) + \phi.
\end{equation}

Formally, $\phi$ is the difference, in dex, between the reduced width of the line of interest evaluated without any blends ($RW_{\mathrm{pure}}$), which corresponds to the abundance measured with the best fit, and the logarithm of the flux subtracted by that same absorption line when the blends are taken into account:

\begin{equation}\label{eq:phi_definition}
    \phi \equiv RW_{\mathrm{pure}} - \log \left \{ \frac{\int_{0}^{+\infty} [ f_{\mathrm{no}}(\lambda') - f_{\mathrm{fit}}(\lambda') ] d\lambda'}{\lambda} \right \},
\end{equation}

\noindent in which $f_{\mathrm{fit}}$ is the best synthetic spectrum fit, and $f_{\mathrm{no}}$ is that best fit with the line of interest removed. The value of $\phi$ tends to zero when line blending is weak. For absorption lines the integral in Eq.~\ref{eq:phi_definition} will always be non-negative. The measurement is flagged as upper limit in cases where $RW_{\mathrm{pure}}$ is lower than $RW_{\mathrm{min}}$. The values of the parameters $\phi$, $RW_{\mathrm{pure}}$, and $RW_{\mathrm{min}}$ for each measurement are listed in the \href{https://zenodo.org/records/14225383}{external appendix}. An exception was made for three stars whose Ir abundances were flagged manually as upper limits. In these cases, the 3513.480 $\AA$ \ion{Co}{I} and the 3513.818 $\AA$ \ion{Fe}{I} lines, which blend with the 3513.640 $\AA$ \ion{Ir}{I} line under evaluation, had their wings strongly under-fitted, likely resulting in an overestimation of these Ir abundances. These stars are marked with one asterisk each in table~4 of the \href{https://zenodo.org/records/14225383}{external appendix}.

Eq.~\ref{eq:threshold2a} can be rewritten in terms of the effective resolution of the spectral line, $R_{\mathrm{eff}} = \lambda/\Delta \lambda$\footnote{In the definition of $R_{\mathrm{eff}}$ all line broadening effects are taken into account in $\Delta \lambda$. That is, $\Delta \lambda$ is the observed resolution element.}:

\begin{equation}\label{eq:threshold2b}
    RW_{\mathrm{min}} = \log 4.5 - \log R_{\mathrm{eff}} - \log S/N + 0.5(\log \delta\lambda - \log \Delta\lambda) + \phi.
\end{equation}

\begin{figure}
	\includegraphics[width=\columnwidth]{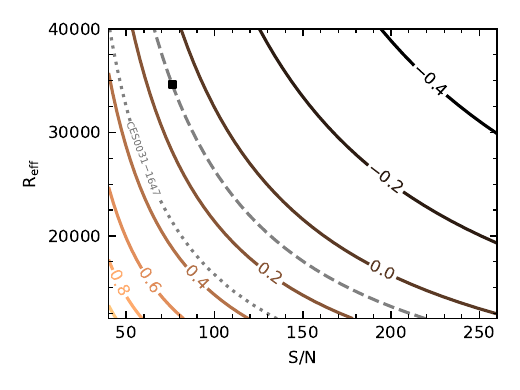}
    \caption{
    Illustration of the method employed for classification of detections and upper limits.
    The solid lines are curves of constant [Hf/Fe]$_{\mathrm{min}}$ calculated for CES\,0031-1647 using Eq.~\ref{eq:threshold2b} and a curve of growth model generated with \texttt{MOOG} for the \ion{Hf}{II} 3399.8 $\AA$ line to convert RW$_{\mathrm{min}}$ to Hf abundances. Their colour scale corresponds to the [Hf/Fe]$_{\mathrm{min}}$ labelled in each solid line.
    The dotted grey line corresponds to the [Hf/Fe] measured in this work, assuming solar Hf abundance of 0.85~dex \citep[][]{Asplund:2009aa}.
    The black square marks the point of the estimated S/N and R$_{\mathrm{eff}}$ of CES\,0031-1647, indicating that the minimum detectable abundance [Hf/Fe]$_{\mathrm{min}}$ for that star with our CERES spectrum is 0.10~dex.
    The unlabelled dashed line crossing that point represents [Hf/Fe]$_{\mathrm{min}}$ = 0.10 for different combinations of S/N and R$_{\mathrm{eff}}$. Abundances to the left side of the dashed line will be detections, while abundances to the right side of the dashed line will result in upper limits.
    }
    \label{fig:threshold}
\end{figure}

An application of Eq.~\ref{eq:threshold2a} (or Eq.~\ref{eq:threshold2b}) takes advantage of the relationship between reduced width and abundance described by the curve of growth theory (CoG). The CoG is a function describing the relationship between the column density of a particular species and the equivalent width of the resulting spectral line, taking into account atomic parameters such as excitation potential, oscillator strength, and the partition function, as well as the model atmosphere. Its usage allows to establish quantitatively a minimum threshold [X/Fe]$_{\mathrm{min}}$ of detectable abundance for a given species in a set of given stellar parameters, with dependence on the quality of the observational data, as exemplified in Fig.~\ref{fig:threshold}. Abundances below this threshold may be regarded as upper limits.

\subsection{Line list}\label{sec:linelist}

The spectral lines measured in this work are shown in Table~\ref{tab:linelist}. The species are \ion{Hf}{II} (1~line), \ion{Os}{I} (2~lines), \ion{Ir}{I}, and \ion{Pt}{I} (1~line). These lines are among the strongest spectral lines for these species in the available wavelength range of our spectra, while also being in relatively clean regions devoid of very strong features (e.g., Balmer lines). For the features under consideration, as well as those blending and in the surrounding wavelengths, we employed \texttt{Linemake}\footnote{\url{https://github.com/vmplacco/linemake}} \citep[][]{Placco_2021} to generate the line lists. The \ion{Hf}{II} line has atomic data from \citet[][]{Lawler_2007}, while for the two \ion{Os}{I} features the oscillator strength values are taken from \citet[][]{Quinet_2006}. The hyperfine and isotopic splitting of the \ion{Ir}{I} line comes from \citet[][]{Cowan:2005aa}, using transition probabilities from \citet[][]{Xu_2007}. For Pt, atomic data was taken from \citet[][]{Hartog_2005}. The abundances used to model blends in the spectra surrounding the Hf, Os, and Ir lines were taken from Papers~I, II, and III. Isotopic fractions were taken from NUBASE2020 \citep[][]{Meija_2016,Kondev_2021}. For \element[][191]{Ir} and \element[][193]{Ir} they are 0.373 and 0.627, respectively. For the isotopes 192, 194, 195, 196, and 198 of Pt, the corresponding fractions are 0.008, 0.329, 0.338, 0.252, and 0.073. For those lines matching the line list published by \citet[][]{Barklem:2000aa}, we adopted constants to model Van der Waals line damping, which were converted from the $\sigma_{ABO}$ values listed in \citeauthor{Barklem:2000aa} to C6 values, using the formula shown in eq.~A.4 from \citet[][]{Coelho_2005}. Otherwise, the default UNSLDc6 approximation from \citet[][]{Unsold:1955aa} was calculated by \texttt{MOOG}. Three auxiliary lines in the 3301 $\AA$ region had their \texttt{Linemake} log~gf values replaced by those available in the Vienna Atomic Line Database (VALD)\footnote{\url{http://vald.astro.uu.se}} \citep[][]{Piskunov_1995,Brooke_2015,K14,K16}. The VALD values generate better fits on the wings of the \ion{Pt}{I} line measured in this study. These lines are also shown in Table~\ref{tab:linelist}.

\begin{table}
    \centering
	\caption{Line list and their respective atomic data.}
	\label{tab:linelist}
    \begin{tabular}{lllll}
    \hline
    Wavelength &          Ion & LEP   & log~gf & Isotope \\
    $[$\AA$]$  &              & [eV]  & [dex]  &         \\
    \hline
    3399.790   & \ion{Hf}{II} & 0.000 & $-$0.570 & --  \\
    3301.565   & \ion{Os}{I}  & 0.000 & $-$0.740 & --  \\
    4420.468   & \ion{Os}{I}  & 0.000 & $-$1.200 & --  \\
    3513.635   & \ion{Ir}{I}  & 0.000 & $-$1.675 & 191 \\
    3513.640   & \ion{Ir}{I}  & 0.000 & $-$1.757 & 191 \\
    3513.643   & \ion{Ir}{I}  & 0.000 & $-$3.080 & 191 \\
    3513.643   & \ion{Ir}{I}  & 0.000 & $-$1.675 & 193 \\
    3513.646   & \ion{Ir}{I}  & 0.000 & $-$1.842 & 191 \\
    3513.647   & \ion{Ir}{I}  & 0.000 & $-$2.959 & 191 \\
    3513.649   & \ion{Ir}{I}  & 0.000 & $-$4.896 & 191 \\
    3513.649   & \ion{Ir}{I}  & 0.000 & $-$1.757 & 193 \\
    3513.651   & \ion{Ir}{I}  & 0.000 & $-$1.927 & 191 \\
    3513.651   & \ion{Ir}{I}  & 0.000 & $-$3.081 & 191 \\
    3513.651   & \ion{Ir}{I}  & 0.000 & $-$4.838 & 191 \\
    3513.652   & \ion{Ir}{I}  & 0.000 & $-$3.080 & 193 \\
    3513.655   & \ion{Ir}{I}  & 0.000 & $-$1.842 & 193 \\
    3513.656   & \ion{Ir}{I}  & 0.000 & $-$2.959 & 193 \\
    3513.659   & \ion{Ir}{I}  & 0.000 & $-$4.888 & 193 \\
    3513.660   & \ion{Ir}{I}  & 0.000 & $-$1.927 & 193 \\
    3513.660   & \ion{Ir}{I}  & 0.000 & $-$3.081 & 193 \\
    3513.661   & \ion{Ir}{I}  & 0.000 & $-$4.821 & 193 \\    
    3301.841   & \ion{Pt}{I}  & 0.814 & $-$1.384 & 195 \\
    3301.842   & \ion{Pt}{I}  & 0.814 & $-$0.907 & 190 \\
    3301.848   & \ion{Pt}{I}  & 0.814 & $-$0.907 & 192 \\
    3301.850   & \ion{Pt}{I}  & 0.814 & $-$2.083 & 195 \\
    3301.855   & \ion{Pt}{I}  & 0.814 & $-$0.907 & 194 \\
    3301.862   & \ion{Pt}{I}  & 0.814 & $-$0.907 & 196 \\
    3301.869   & \ion{Pt}{I}  & 0.814 & $-$1.129 & 195 \\
    3301.870   & \ion{Pt}{I}  & 0.814 & $-$0.907 & 198 \\
    \hline
    \multicolumn{5}{l}{Auxiliary lines with changes in log~gf:\tablefootmark{a}}
    \\ \hline
    3301.673   & \ion{Ti}{II} & 1.164 & $-$1.842 & --  \\
    3301.913   & \ion{Fe}{I}  & 3.234 & $-$1.748 & --  \\
    3302.095   & \ion{Ti}{II} & 0.151 & $-$2.165 & --  \\
    \hline
    \end{tabular}
    \tablefoot{LEP is the lower excitation potential. The last column shows the isotopes for the line in which isotopic splitting was taken into consideration. See text for references. \\
    \tablefoottext{a}{Blending lines in which we adopted log~gf values different from those compiled in \texttt{Linemake}.}}
\end{table}

In the particular case of the 3399.79 $\AA$ \ion{Hf}{II} line, there is a blend with a strong NH line ($\lambda$=3399.80 $\AA$, lower excitation potential LEP=0.264~eV, log~gf=$-$1.058) that requires very careful consideration. Its log~gf value adopted in this study is from the \texttt{Linemake} database, the same value from the line list compiled by \citet[][]{Kurucz:1995aa}\footnote{\url{http://kurucz.harvard.edu/molecules/old/nh.asc}}. However, it is important to note that the figure found for its oscillator strength in VALD \citep[][]{Fernando_2018} is 0.3~dex lower than the NH log~gf values from \texttt{Linemake}, which were used as reference in the measurements of the 3360 $\AA$ NH band in Paper~II, thus we adopted them in the 3399 $\AA$ region for consistency, since the adopted N abundances come from Paper~II as well. The log~gf values listed in VALD also differ from \texttt{Linemake} by roughly $-$0.3~dex for most NH lines in the 3360 $\AA$ band. For comparison, the log~gf values for the NH band in \citet[][]{Spite:2005aa} were also adopted from the Kurucz line list. In Appendix~\ref{sec:modelNH} a short discussion on the theoretical modelling of oscillator strength values for NH lines is done.

\begin{figure}
	\includegraphics[width=\columnwidth]{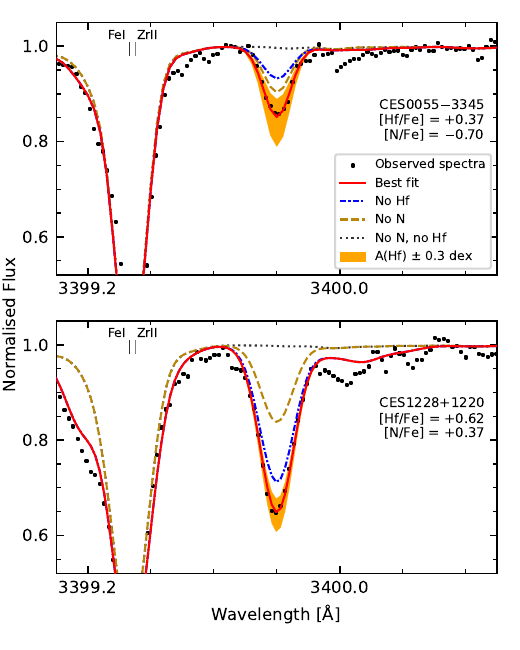}
    \caption{
    Portion of spectra surrounding the \ion{Hf}{II} line at 3399.8 $\AA$ for two stars in our sample.
    Black dots represent the observed spectra.
    The red solid lines represent the best fit, with their respective [Hf/Fe] shown on the right.
    Blue dash-dotted lines show synthetic spectra without Hf.
    Golden dashed lines represent synthetic spectra when N is removed.
    Grey dotted lines correspond to the synthetic spectra without neither N nor Hf.
    Orange shaded areas display a $\pm$~0.3~dex interval around the best fit.
    The N abundances shown are from Paper~II.
    }
    \label{fig:spec_3399}
\end{figure}

Accurate N measurements are crucial for accurate Hf abundances using the line chosen in this work. After testing synthetic spectra generated with stellar parameters typical of our sample (T$_{\mathrm{eff}}$=5000~K, log~g=2.0, [Fe/H]=-2.5, $v_t$=2.0 km s$^{-1}$) we found that the equivalent width of the (isolated) NH line is almost four times larger than the equivalent width of the (also isolated) \ion{Hf}{II} line in the linear part of the curve of growth for solar [Hf/N]. In the most evolved stars in our sample, the pollution of the surface with dredged-up N may result in the NH line dominating the blend entirely, unless the star also displays a large Hf enhancement. Fig.~\ref{fig:spec_3399} shows an example of the contributions of both species for the 3399.8 $\AA$ feature. The top (bottom) panel of Fig.~\ref{fig:spec_3399} shows a star with a low (high) $\phi$ value, defined in Eq.~\ref{eq:phi_definition}, i.e. a weak (strong) contribution from the NH line for the blend with Hf.

\subsection{Uncertainties}

The uncertainties for A(X) values were calculated taking into account both the process of line formation, by estimating their sensitivities to the model atmospheres, and observational uncertainties. For a measurement of A(X), the total uncertainty is composed by the sensitivities of A(X) towards the internal uncertainties of the atmosphere models listed in Paper~I and the sensitivities of A(X) towards uncertainties in neighbouring blends. From the observational part, we take into account the uncertainty in the continuum placement, defined as a function of S/N, as well as the residuals of the fit of the synthetic spectrum on the observed data, weighted by their distance to the central wavelength of the line of interest. Formally, the uncertainties $\sigma_{\mathrm{A(X)}}$ for each detection of A(X) are:

\begin{equation}\label{eq:propagation}
    \sigma_{\mathrm{A(X)}} = \sqrt{\sigma_{\mathrm{\Delta}}^2 + \sigma_{\mathrm{c}}^2 + \sigma_{\mathrm{f}}^2},
\end{equation}

\noindent where $\sigma_{\mathrm{\Delta}}^2$ is the quadrature of the sensitivities of A(X) to the uncertainties of the atmospheric parameters $\overrightarrow{a} =$ (T$_{\mathrm{eff}}$, log~g, [Fe/H], $v_t$) in a particular star. The sensitivity of the abundances to each atmospheric parameter a$_i$ is estimated by measuring how much A(X) change for an 1$-\sigma$ variation of that atmospheric parameter a$_i$, while keeping the other parameters fixed. The adopted uncertainties of the atmospheric parameters were derived in Paper~I and their values are $\overrightarrow{\sigma_{a}} =$ (100~K, 0.04~dex, 0.13~dex, 0.5~km s$^{-1}$).

The uncertainty in continuum placement $\sigma_{\mathrm{c}}$ is the change in A(X) that results from the uncertainty generated by noise in the continuum -- the choice of the continuum placement may result in a synthetic line that is shallower or deeper than the true spectral line. We estimated $\sigma_{\mathrm{c}}$ as the change in A(X) corresponding to the change in the reduced width $\Delta RW$ of the spectral line under measurement, which results from a change $\Delta c$ in the position of the (normalised) spectral continuum. Based on the line depth of $\sigma_{W_{\lambda}}$, calculated with Eq.~\ref{eq:cayrel} with typical values of $\delta\lambda = 0.014$ $\AA$ and $\Delta\lambda = 0.1$ $\AA$ from our spectra, $\Delta c$ is defined as 0.5(S/N)$^{-1}$, with S/N measured near the spectral line of interest. Under the assumption that the line has a Gaussian profile, we estimated $\Delta RW$ as:

\begin{equation}\label{eq:deltaRW}
    \Delta RW = [2\;\ln(10)\;l\;(S/N)]^{-1},
\end{equation}

\noindent by taking the derivative of RW with respect to the depth of the normalised spectral line $l$.

To estimate the uncertainty of the fit $\sigma_{\mathrm{f}}$ we applied a weight $W_{\epsilon}$ on the residuals of the fit $\epsilon(\lambda) = |f_o(\lambda) - f_s(\lambda)|$, where $f_o(\lambda)$ and $f_s(\lambda)$ are the fluxes of the observed and synthetic spectra in a given wavelength, respectively. The weight:

\begin{equation}\label{eq:weight}
    W_{\epsilon}(\lambda) = \exp{\left [ -\frac{(\lambda-\lambda_{c})^{2}}{2\sigma^{2}} \right ]}
\end{equation}

\noindent is intended to discard any bias from residuals far from the core of the line. In the above equation, $\sigma$ corresponds to the width of the line under evaluation, and $\lambda_{c}$ is its central wavelength. We then fitted a Gaussian in the weighted residuals $\epsilon_{W}(\lambda) = \epsilon(\lambda)W_{\epsilon}(\lambda)$, and used the value of the fitted Gaussian at the central wavelength of the spectral line as the total change in the line depth due to uncertainty in the fit. Since this work deals with blended lines, we estimated the fraction of the subtracted flux in the absorption line corresponding to the species of interest (Hf, Os, Ir, or Pt) in each star, and from that fraction we calculated the corresponding change in line depth -- i.e., the total change in line depth times the fraction of the subtracted flux coming from the species being evaluated -- due to the fit uncertainty for each species. That change in line depth was used to estimate the change in abundance $\sigma_{\mathrm{f}}$.

\begin{figure}
	\includegraphics[width=\columnwidth]{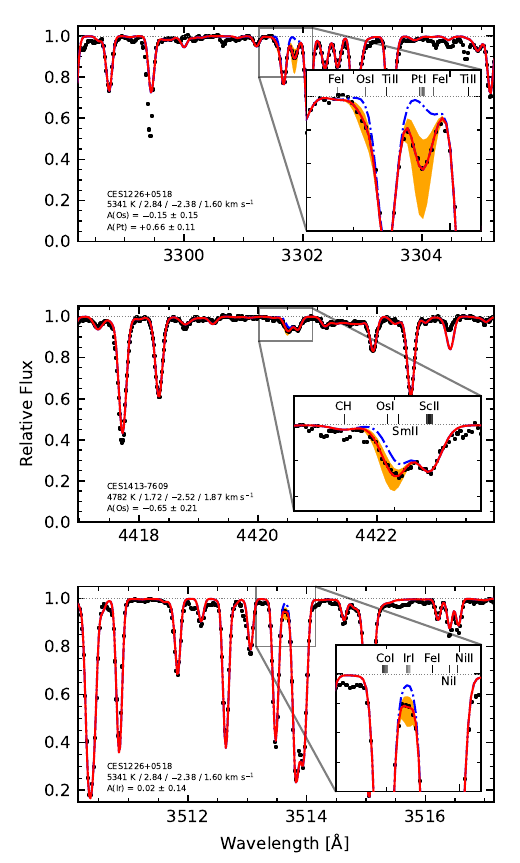}
    \caption{
    Examples of spectral fit in Os, Ir, and Pt lines.
    Black markers: observed spectra.
    Red solid line: best fit.
    Blue dash-dotted line: spectrum with the species of interest removed.
    Orange shaded areas: $\pm$~0.3~dex interval around the best fit.
    Species creating the strongest surrounding lines are labelled in the insets, with the hyperfine/isotopic splitting marked when considered in the line list. The darkest shades of grey indicate the strongest lines.
    }
    \label{fig:fit_osir}
\end{figure}

The uncertainties for each measurement, as well as their respective sensitivities, and partial uncertainties $\sigma_{\mathrm{c}}$ and $\sigma_{\mathrm{f}}$ are shown in the \href{https://zenodo.org/records/14225383}{external appendix}. The uncertainties of the abundance ratios discussed in this work were calculated from the A(X) uncertainties, using equations A19 and A20 from \citet[][]{McWilliam:1995aa}. When abundance ratios are in square bracket notation, solar abundances and their respective uncertainties are from \citet[][]{Asplund:2009aa}\footnote{Except the solar abundance of Fe [A(Fe)$_{\sun} = 7.52$], which, as in Paper~I, comes from \citet[][]{Caffau_2010}.}. In that case, the uncertainties of solar abundances were also added in quadrature when propagating the uncertainties of [A/B].

\section{Results and discussion}\label{sec:results}

\begin{figure}
	\includegraphics[width=\columnwidth]{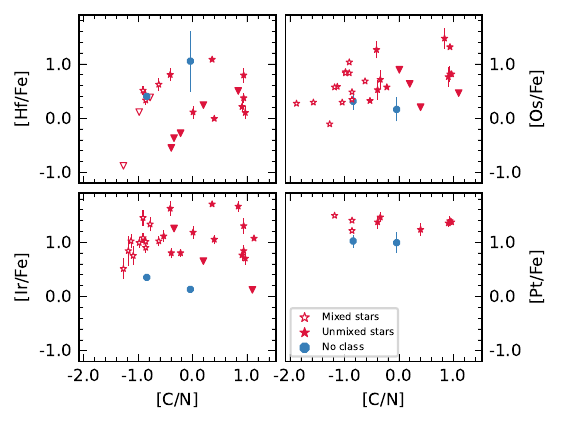}
    \caption{
    [X/Fe] versus [C/N] ratio from Paper II.
    Triangles represent upper limits. Open/filled triangles follow the same classification of star markers.}
    \label{fig:x_cn}
\end{figure}

The calculated abundances for Hf, Os, Ir, and Pt, as well as the stellar parameters from Paper~I are shown in Table~\ref{tab:abundances}. Out of our sample of 52 red giants, we derived \ion{Hf}{II} abundances for 19 of them, \ion{Os}{I} for 33 stars, \ion{Ir}{I} for 32 of the targets, and \ion{Pt}{I} in 18 stars, upper limits excluded. Examples of spectral fitting for the elements of the third r-process peak are shown in Fig.~\ref{fig:fit_osir}. Abundances derived in Papers I, II, and III were used to synthesise the neighbouring lines. They were assumed solar-scaled in cases of species without abundance information in previous CERES papers. Eight of the objects ended without any measured abundances for these four species due to either insufficient S/N in their spectra and/or glitches in the spectra at the wavelengths of interest. We observed that the extra-mixing that takes place after the RGB Bump does not have influence on the abundances of these elements. As shown in Fig.~\ref{fig:x_cn}, apart from the distinction between mixed and unmixed stars around [C/N]\,$\sim$\,0.5\,dex that was used in Paper\,II as indicator of extra mixing, we could not find clues that the stars in the upper RGB have an overabundance of Hf, Os, Ir, and Pt with regards to the stars in the lower part of the RGB.

\subsection{Hafnium}

\begin{figure}
	\includegraphics[width=\columnwidth]{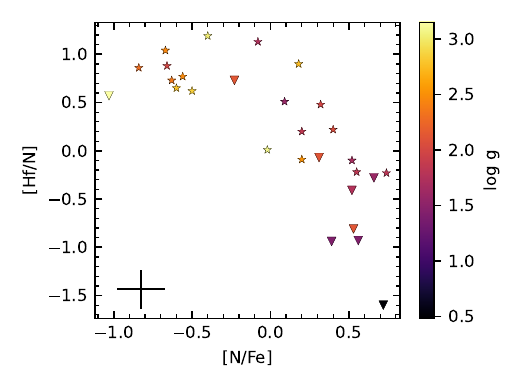}
    \caption{
    Ratio between Hf and N as function of the N abundances from Paper~II, coloured by surface gravity.
    Triangles are upper limits.
    The error bars on the lower left represent the median uncertainties.
    }
    \label{fig:hfn_nfe}
\end{figure}

As previously discussed in Sect.~\ref{sec:chem}, the strong blend between \ion{Hf}{II} and a NH line requires the N abundances derived in Paper~II to be considered in the synthesis of the 3399.8 $\AA$ line. In Fig.~\ref{fig:hfn_nfe} it is shown that the [Hf/N] ratio anti-correlates with [N/Fe] -- Pearson and Spearman correlation coefficients $\rho \approx$ $-$0.7, with p-values $\approx$ 10$^{-4}$. Fig.~\ref{fig:hfn_nfe} is colour-coded by surface gravity to trace the positions of the stars in the red giant branch\footnote{Except for CES\,2250$-$4057, which is assumed to be a horizontal branch star, see Paper~II.}. The larger concentration of N-rich objects among the stars with upper limits in Hf suggests that the use of the 3399 $\AA$ line for Hf measurements is more effective when [N/Fe]~$\lesssim$~0.1~dex. Regardless, Hf can be detected in some N-rich stars, as indicated by a few measurements in stars with [N/Fe] larger than 0.4~dex.

\begin{figure*}
    \includegraphics[width=\textwidth]{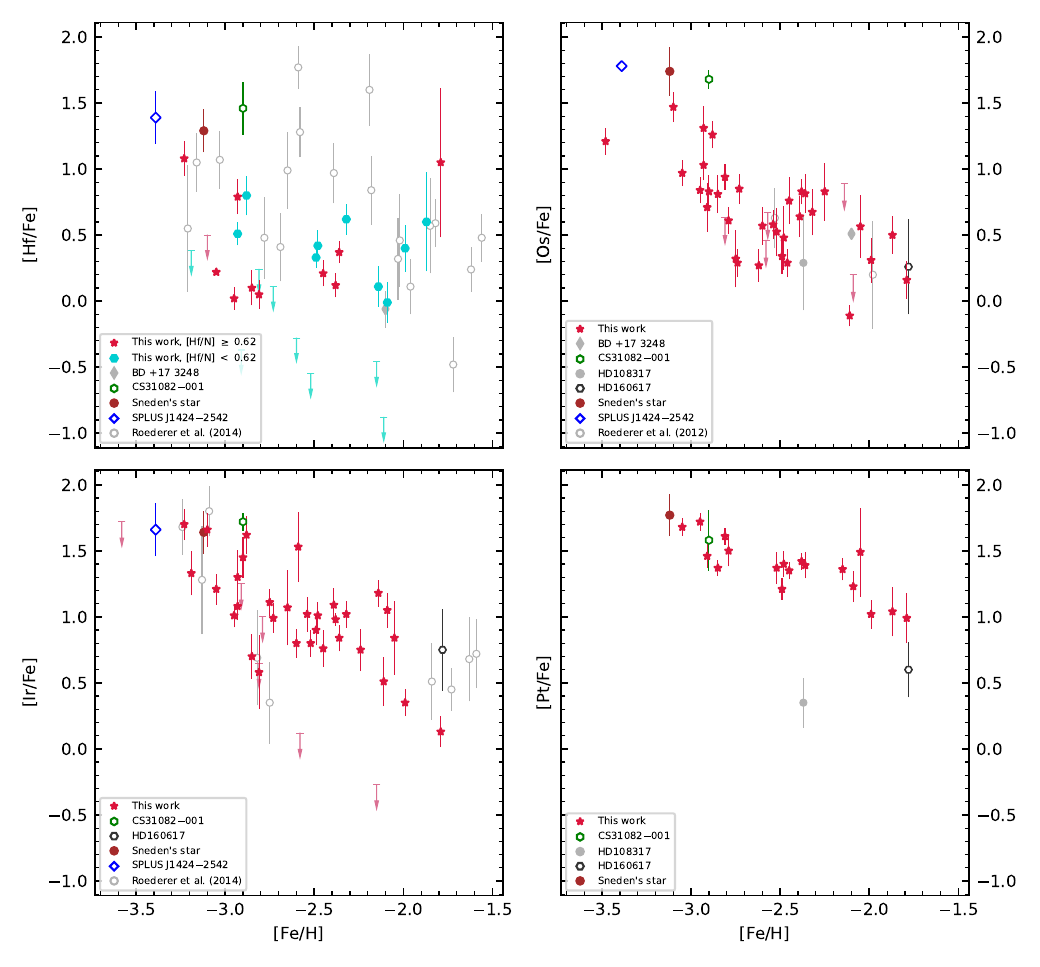}
    \caption{
    [Hf,Os,Ir,Pt/Fe] plotted against metallicity [Fe/H], with BD\,$+$17\,3248 \citep[][]{Roederer_2010}, CS\,22892$-$052 \citep[Sneden's star,][]{Sneden:2003aa}, CS\,31082$-$001 (\citet[][]{Hill:2002aa} for Hf, \citet[][]{Barbuy_2011} for Os, Ir, Pt), HD\,108317 \citep[][]{Roederer_2014b}, HD\,160617 \citep[][]{Roederer:2012ac}, SPLUS\,J1424$-$2542 \citep[][]{Placco_2023}, and data from \citet[][]{Roederer:2012ab} and \citet[][]{Roederer:2014ac} for comparison. 
    Arrows are upper limits.
    The data from \citet[][]{Roederer:2014ac} also includes abundances for CS\,22892$-$052, CS\,31082$-$001, and HD\,108317.
    }
    \label{fig:xfe_feh}
\end{figure*}

Due to the heavy blending between the Hf line and a NH line and the anti-correlation seen in Fig.~\ref{fig:hfn_nfe}, it is important to verify that the enhancement of NH does not lead to an underestimation of Hf due to the strong blend. A sanity check is evaluating the distribution of [Hf/Fe], which, in case of an eventual bias introduced by the NH blend, would be different for N-rich and N-poor stars. We divided our 19 results for [Hf/Fe] in two subsets, [Hf/N]-rich/poor, using the [Hf/N] median of 0.62~dex as delimiter. The two subsets are displayed in Fig.~\ref{fig:xfe_feh}, where they do not seem to differ by visual inspection. A two-sample Kolmogorov–Smirnov (KS) test to compare these subsets also does not identify any differences. Using the method \texttt{stats.ks\_2samp} from the \texttt{Python} library \texttt{Scipy} \citep[][]{scipy:2020} to evaluate the KS statistic we found a value of 0.38, too low to reject the null hypothesis that these two samples could have been drawn from different distributions. Furthermore, when we compare our results with those from \citet[][]{Roederer:2014ac}, also shown in Fig.~\ref{fig:xfe_feh}, while a difference in the median [Hf/Fe] is clear, the spread in Hf abundances is similar in both studies. The median absolute deviations\footnote{In this work, the scale factor of the median absolute deviation was adjusted for a normal distribution.} of [Hf/Fe] are 0.4~dex for the CERES sample and 0.5~dex in \citet{Roederer:2014ac}.

\begin{figure*}
    \includegraphics[width=\textwidth]{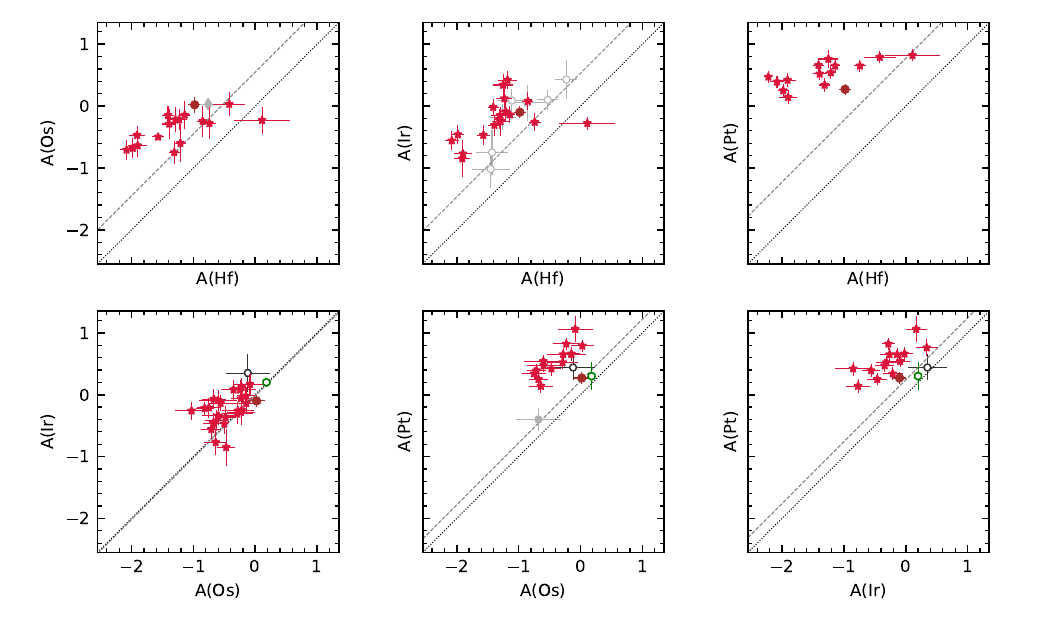}
    \caption{
    The four species under study plotted against each other in terms of A(X).
    Markers are the same as in Fig.~\ref{fig:xfe_feh}.
    Dotted lines represent the identity function.
    Dashed lines represent solar-scaled values -- i.e., their y-intercept give the A(y-axis)/A(x-axis) solar values.
    Upper limits are omitted in this figure.
    The x-axis values are the same in all three panels. 
    }
    \label{fig:ax_vs_ax}
\end{figure*}

In summary, our Hf abundances are not affected by the blend with the NH line, and we attribute the anti-correlation seen in Fig.~\ref{fig:hfn_nfe} to the lack of stars with extreme enhancements in Hf which, if existed, could occupy the upper right area of the plot. Regardless, future studies intending to use the 3399.8 $\AA$ \ion{Hf}{II} line must take N abundances into account for optimisation of the target selection procedure.

\subsection{Osmium}

In this work we derive 33 \ion{Os}{I} abundances plus 5 upper limit values. Our results show that Os abundances display enhancement at lower metallicities, see Fig.~\ref{fig:xfe_feh}. In the CERES sample, [Os/Fe] has an anti-correlation with [Fe/H]: their Pearson coefficient is $-$0.66, p-value $\approx$ 10$^{-5}$. This means [Os/Fe] increases with decreasing metallicities, stopping around the metallicity of CS\,22892$-$052 \citep[][hereafter Sneden's star]{Sneden:2003aa}. We analyse two lines for this element, located at 3301.565 $\AA$ and 4420.468 $\AA$ (see Table~\ref{tab:linelist} for their atomic parameters). In stars with measurements of both spectral lines, the bluest line has a median abundance 0.11~dex larger than the reddest one, and the median absolute deviation of the line-by-line difference is 0.13~dex. Such difference between the blue and the red \ion{Os}{I} lines has the same magnitude of the effect from Rayleigh scattering found by \citet[][]{Reichert_2020} in Sr. The A(Os) = $+$0.02 adopted by \citet[][]{Sneden:2003aa} for Sneden's star is from the line at 3301.565~\AA. Comparing our abundances with the measurement in Sneden's star suggests a relative enhancement in Os in the latter. Other measurements available in the literature show an agreement with our [Os/Fe] trend, as seen in Fig.~\ref{fig:xfe_feh}. Data on Os abundances has been so far very scarce in the literature, preventing a more comprehensive comparison with previous studies.

\subsection{Iridium}

We provide \ion{Ir}{I} abundances or upper limits for 38 red giants, of which 32 are detections. Similarly to Os, in Fig.~\ref{fig:xfe_feh} our results show an anti-correlation between [Ir/Fe] and [Fe/H] in the metallicity range of our sample. The Pearson correlation coefficient is $-$0.68, with p-value $\approx$ 10$^{-5}$. Interestingly, this anti-correlation fills a gap in the data from \citet[][]{Roederer:2014ac}. In their results, the two stars in the -3 $<$ [Fe/H] $<$ -2 interval are Ir-deficient compared to those in our main anti-correlation pattern. At the same time, our results also present some Ir-deficient stars within the same metallicity range. \citet[][]{Sneden:2003aa} adopted A(Ir) = $-$0.10 for Sneden's star. However, that value is an average of two lines: if we consider only their measurement in the same spectral line adopted in this work (3513~\AA), the A(Ir) of Sneden's star increases by 0.05~dex, placing it on the identity function in the bottom panel of Fig.~\ref{fig:ax_vs_ax}, considering the uncertainties. This re-scaling makes their result qualitatively similar to one third of our sample, whose error bars cross the identity function in the A(Ir) vs. A(Os) panel in Fig.~\ref{fig:ax_vs_ax}.

\subsection{Platinum}
\label{sec:pt}

Abundances of \ion{Pt}{I} are calculated for 18 stars, all of them classified as detections. Like in the other two species of the third r-process peak, [Pt/Fe] displays a negative slope against metallicity, albeit milder. In spite of that, the Pearson correlation coefficient between both quantities is higher, $-$0.81, with p-value $\approx$ 10$^{-5}$, being $-$0.89 if the outlier CES\,0547$-$1739 ([Pt/Fe] = 1.49 $\pm$ 0.33) is not considered.

Compared to data points available in the literature, our results are consistent with Sneden's star \citep[][]{Sneden:2003aa} and CS\,31082$-$001 \citep[][]{Barbuy_2011} in the abundance space shown in Fig.~\ref{fig:xfe_feh}. In their reanalysis of CS\,31082$-$001, \citet[][]{Ernandes_2023} adopts a value of A(Pt) which is 0.3 dex lower than that found by \citeauthor{Barbuy_2011}, still consistent with the trend found in the CERES sample. Meanwhile, the results for HD\,108317 \citep[][]{Roederer_2014b} and HD\,160617 \citep[][]{Roederer:2012ac} display deficiency of Pt compared to our general trend. Apart from \citet[][]{Sneden:2003aa} and \citet[][]{Ernandes_2023}, all the other works relied on spectral lines below 3000~\AA, observed from space with the Hubble Space Telescope. \citeauthor{Sneden:2003aa} used both space- and ground based observations, including the 3301 $\AA$ \ion{Pt}{I} line adopted in this work. Their Pt abundance for Sneden's star based on that individual line is 0.06~dex larger than their adopted value, while the log~gf adopted by \citeauthor{Sneden:2003aa} is 0.14 dex larger than the log~gf used in our measurements. Hence, a more accurate comparison with our results would require a shift of $+$0.2~dex in the A(Pt) = 0.27~dex abundance adopted by \citet[][]{Sneden:2003aa}, assuming that the systematics generated by the different model atmospheres employed are negligible. That shift in the Pt abundance would put Sneden's star in better agreement with our results in Fig.~\ref{fig:ax_vs_ax}.

\subsection{Abundance trends}

In Figs.~\ref{fig:xfe_feh} and \ref{fig:ax_vs_ax} it is clear that the Os, Ir, and Pt abundances have a similar behaviour. They are all anti-correlated with metallicity and show similar slopes in the interval covered by our sample. When compared to Hf, the first post-lanthanide element, they also seem to present the same pattern: a steady increase in A(Os,Ir,Pt) against Hf up to A(Hf) $\sim$ $-$1.0, followed by an apparent plateau in the more metal-rich stars in Fig.~\ref{fig:ax_vs_ax}. It is not clear if there is such plateau in Ir vs. Hf, because Ir measurements are lacking for stars between $-$0.8~$<$~A(Hf)~$<$~0.0~dex. A larger sample in the $-$2~$<$~[Fe/H]~$<$~$-$1 interval could enable one to tell if that plateau is real or just a selection effect. Hafnium is the only element analysed in this study with a relatively large share of production by the s-process in the Solar System \citep[49\%,][]{Simmerer:2004aa}, thus, that plateau, if real, may indicate an increase of the s-process contribution to the metal enrichment of that group of stars.

\subsubsection{The Eu-poor tail}
\label{sec:eutail}

\begin{figure}
    \includegraphics[width=\columnwidth]{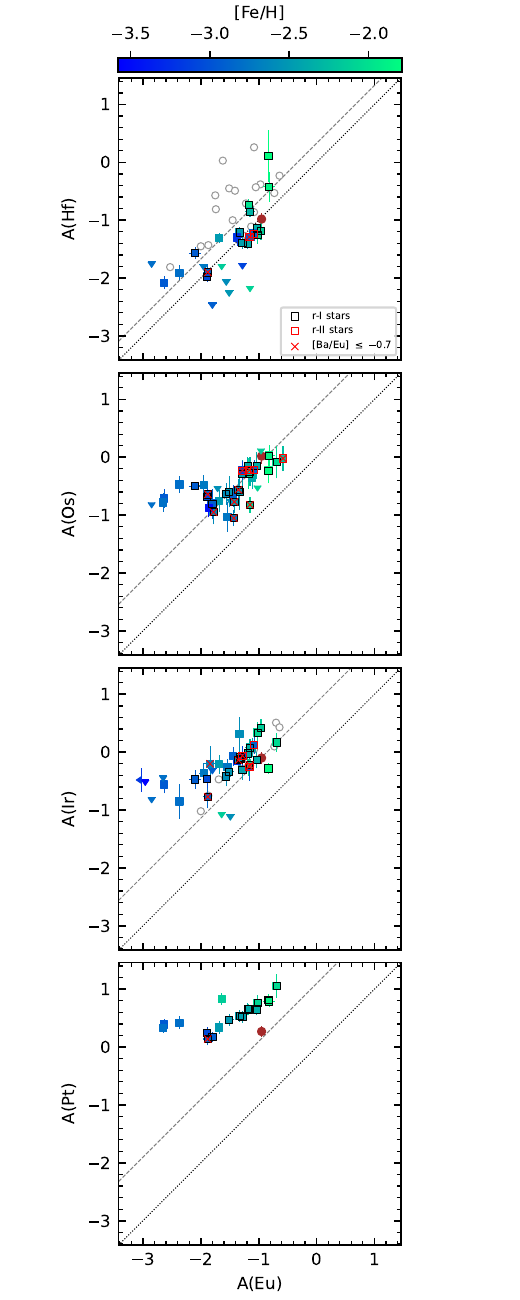}
    \caption{
    A(Hf,Os,Ir,Pt), from this work, compared to A(Eu) (Paper~III).
    Markers are coloured by metallicity.
    Triangles pointing to the left represent upper limits in A(Eu) and detections in the element represented in the y-axis.
    Triangles pointing down show upper limits for either Hf, Os, or Ir, and detections for Eu.
    All other markers follow Fig.~\ref{fig:xfe_feh}.
    Lines follow Fig.~\ref{fig:ax_vs_ax}.
    [Ba/Eu] values are from Paper~III.
    }
    \label{fig:ax_vs_eu}
\end{figure}

In Fig.~\ref{fig:ax_vs_eu} the abundances of the four elements analysed in this work are compared to Eu abundances derived in Paper~III. Fig.~\ref{fig:ax_vs_eu} suggests the existence of a tail in the Eu-poor end for Os, Ir, and Pt. The Eu-tails consist of plateaus of the elements from the third r-process peak for stars with A(Eu) $\lesssim$~$-$1.8~dex. These plateaus seem roughly centred on A(Os)~$\sim$~$-$0.7, A(Ir)~$\sim$~$-$0.6, and A(Pt)~$\sim$~$+$0.25~dex. The data are coloured by metallicity, and it can be seen that the more metal-rich stars in our sample are absent in the Eu-poor tails, making this a trend that only occurs at the low metallicity end of our sample. It is interesting to note that the more metal-poor objects where Os and Ir follow Eu are r-II stars, using the definition given by \citet[][]{Christlieb_2004} -- i.e., [Eu/Fe]~$>$~$+$1.0 and [Ba/Eu]~$<$~0 -- highlighted by red boxes in Fig.~\ref{fig:ax_vs_eu}. Meanwhile, r-I stars are marked by black boxes, while r-poor stars are unmarked. These more metal-poor stars are part of the large spread in Eu which appears in [Fe/H]~<~$-$2.5, known in the literature for a long time \citep[see, e.g., the review from][and references therein for a discussion]{Sneden:2008aa}.

The results shown in Fig.~\ref{fig:ax_vs_eu} suggest that the elements in the third r-process peak have a different behaviour with respect to Eu in the [Fe/H]~<~$-$2.5 region, despite all these four species (Eu, Os, Ir, Pt) being nearly pure r-process elements \citep[][]{Simmerer:2004aa,Bisterzo:2014aa}. In a scenario where these elements are co-produced with Eu, a simple linear relation should appear in their respective panels in Fig.~\ref{fig:ax_vs_eu}. Since this linear relation breaks for Eu-poor stars, models must take into account some particular scenarios where elements of the third r-process peak have (relatively) larger production rates compared to lanthanides. Interestingly, Hf is the only one among the four elements analysed here that seems to follow Eu, despite being (by a large margin) the most s-process rich species (in the Solar System) studied in this work.

One caveat to the behaviour of the third r-process peak elements against Eu is the possibility of metallicity-dependent LTE effects in Os, Ir, and Pt abundances, which are still yet to be computed. However, such corrections should be similar in all the three (Os, Ir, Pt) species, making LTE effects as the cause of the Eu-tail unlikely. Another caveat is of observational nature: the existence of a large scatter below the A(Os,Ir)~$<$~$-$1.0 threshold may be currently invisible due to the corresponding spectral lines being too weak. If such star-to-star scatter in third-peak elements exists, S/N values larger than those available in our spectra would be necessary to observe it for Os and Ir. Nevertheless, the lack of scatter of Pt abundances in the Eu-poor region suggests that the Eu-tail must be robust, because the Pt line at 3301 $\AA$ is stronger than those lines used for Os and Ir in this work, and the influence of blends in the Pt line is mild to negligible, as indicated by the very low values of $\phi$ in table~5 of the \href{https://zenodo.org/records/14225383}{external appendix}.

Interestingly, a pattern in the Eu abundances with similar shape and similar metallicity compared to that from CERES has been noticed in the work of \citet[][]{Krishnaswamy_Gilroy_1988}. In their fig.~14 they compare A(Eu) with an average of A(Sr,Y,Zr), resulting in a similar trend: the average of A(Sr,Y,Zr) grows mildly in the more Eu-poor stars, and starts to increase more rapidly in the more Eu-rich stars in their sample. That change in slope was attributed to the onset of the s-process. However, in the case of Os, Ir, and Pt, whose s-process contributions in the Solar System are lower than 10\% (like Eu), that suggestion is unlikely to hold. \citet[][]{Hansen_2014ab} also measured a similar trend of Eu with respect to Mo and Ru, which have a large s-process production share in the Solar System compared to the r-process third peak elements. In that case (i.e., Mo, Ru), the Eu-tail also forms below A(Eu)~$\approx$~$-$1.8~dex. In CERES data, we have also found a similar tail when the third r-process peak is compared to Ba (see Fig.~\ref{fig:osba}).

When putting stellar abundances into the context of Galactic Chemical Evolution, it is important to consider mono-enrichment. That is, if the star was formed from a gas previously enriched by a single event of stellar nucleosynthesis. \citet[][]{Hartwig:2018aa} proposed the Mg/C ratio as a tracer for mono-enrichment, based on Population~III supernovae yields. From the [Mg/C] ratios derived in Paper~II, no star in our sample at first glance appears mono-enriched. However, this could be due to our sample selection removing Carbon Enhanced Metal-Poor (CEMP) stars from this study, and, also, dilution must be taken into consideration when comparing models with observations \citep[see][]{Hansen_2020,Magg_2020}. Still, the shape observed in Fig.~\ref{fig:ax_vs_eu}, with a transition from a flat third peak to co-production with Eu around A(Eu) $\approx$ $-$1.8, if real, may mean some form of transition in the chemical enrichment of the Universe. In order to look for traces of the s-process in the low metallicity range in which the Eu tail appears, we singled out two members of the metal-poor, Eu-poor tail (CES\,1322$-$1355, CES\,1402$+$0941), as well as two metal-poor, Eu-rich, Os-rich stars (CES\,2254$-$4209, CES\,2330$-$5626), for measurement of Pb. However, the strongest Pb feature in our spectral range, at 4057~\AA, is undetectable in the spectra of any of the test targets. An alternative to the measurements of Pb would be performing measurements of isotopic ratios of Ba to trace the s-process, \citep[as in, e.g.,][]{Magain_1993,Mashonkina_2006,Gallagher_2010,Gallagher_2012,Gallagher_2015,Mashonkina_2019}, ideally with a full 3D, non-LTE treatment like done in the Sun by \citet[][]{Gallagher_2020}. Such a Ba analysis could yield valuable information on the contribution to the s-process, if present, to the enrichment of the interstellar medium before the formation of these stars. However, such analysis requires S/N values much larger than those available in our sample. If the contribution from the s-process could be ruled out (i.e., if at least some of these stars are r-process pure), modelling of previous enrichment events would be facilitated. In Paper~III the stars with [Ba/Eu]~$\leq$~$-$0.7 were discussed as being r-pure stars, i.e., without s-process contribution. Interestingly, none of those appear in the Eu-tail in Fig.~\ref{fig:ax_vs_eu}, while one is at the transition zone around A(Eu)~$\approx$~$-$1.9~dex.

\subsubsection{Shape of the third r-process peak}
\label{sec:shape}

\begin{figure}
    \includegraphics[width=\columnwidth]{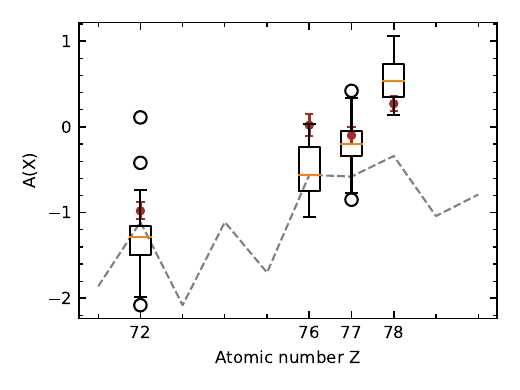}
    \caption{
    Box plots representing the distributions of A(Hf,Os,Ir,Pt) in our sample. The orange lines represent the medians.
    The grey dashed line is the solar-scaled abundance, normalised to the median of the Os abundances from this work.
    Open black circles are the outliers in our sample.
    The filled brown circles are Sneden's star abundances as published by \citet[][]{Sneden:2003aa}.}
    \label{fig:pattern}
\end{figure}

The Ir abundances in this work are, on average, 0.18~dex ($\approx$~1-$\sigma$ in abundances) larger than Os, with the standard deviation of the difference being 0.17~dex. Hence, our results show Ir (Z=77) and Os (Z=76) breaking the even/odd effect in stellar abundances, indicating a monotonic increase in abundances with respect to atomic number from Os to Pt. This increase can be seen in Fig.~\ref{fig:pattern}, which compares CERES data with Sneden's star and solar-scaled abundances. The monotonic increase in abundances can also be visualised, for instance, in the panels of Fig.~\ref{fig:ax_vs_eu}, with the identity function line crossing the data points in the top panel (Hf vs. Eu), and with the data points getting progressively further away from the identity function in the panels containing Os, Ir and Pt. This indicates an enhancement of the third peak with respect to Eu.

The lower left panel of Fig.~\ref{fig:ax_vs_ax} (Ir vs. Os) shows that about a third of our sample with detections for both species has a "flat" A(X) vs. atomic number profile, i.e., their Os and Ir abundances are similar if we take the error bars into account. That is the case, for instance, in the Sun \citep[][]{Asplund:2009aa,Lodders:2009aa}, in CS\,31082$-$001 \citep[][]{Barbuy_2011}, in HD\,222925 \citep[][]{Roederer_2018,Roederer_2022}, and in Sneden's star \citep[][]{Sneden:2003aa}. On the other hand, HD\,160617 \citep[][]{Roederer:2012ac}, displays an Ir/Os ratio similar to the stars in our sample on the opposite end, i.e., those whose Ir abundance is larger than the Os abundance by more than 1-$\sigma$. Two thirds of the stars in the CERES sample have significantly larger abundances of Ir than Os. In the case of Pt, if the abundance from \citet[][]{Sneden:2003aa} is shifted by $+$0.2~dex as discussed in Sect.~\ref{sec:pt}, then the A(Pt) from Sneden's star would differ by less than 1-$\sigma$ from the median A(Pt) shown in Fig.~\ref{fig:pattern}.

\subsection{Comparison to calculations: new r-process?}\label{sec:models}

\begin{figure}
    \includegraphics[width=\columnwidth]{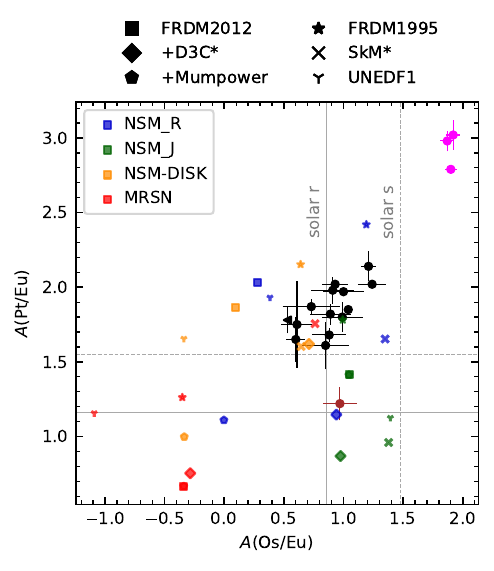}
    \caption{
    A(Pt/Eu) vs A(Os/Eu) for observations and nucleosynthesis calculations. Black circles with error bars are abundance detections derived in this work, while left-facing triangles are upper limits in Os. Sneden's star is depicted in brown, while the solar r- (s-) residuals are given by the grey solid (dashed) lines. The calculations assume different r-process sites (colours) and different nuclear inputs (symbols). None of the calculations are able to reproduce the large A(Pt/Eu)- and A(Os/Eu)-values of the three stars in the upper right (magenta), which make up the Eu-tail in Fig.~\ref{fig:ax_vs_eu}.}
    \label{fig:obsVsCalc}
\end{figure}

In this section we compare the abundances derived in this work to r-process nucleosynthesis calculations using the nuclear reaction network \texttt{WinNet} \citep[][]{Reichert_2023} and representative trajectories of different r-process sites: neutron star mergers (NSM\_R from \citealt{Rosswog_2013} and NSM\_J from \citealt{Jacobi_2023}), their disks (NSM-DISK from \citealt{Wu_2016}), and magneto-rotational supernovae (MRSN from \citealt{Reichert_2021}). Fig.~\ref{fig:obsVsCalc} shows the ratios of Pt/Eu and Os/Eu for the observed abundances and for our calculations. The astrophysical variability is given by the different scenarios (shown with different colours). For each of them a variety of different nuclear physics inputs is explored (shown with different markers). We have used the FRDM2012 mass model \citep[][]{Moller_2016} and beta decay rates \citep[][]{Moller_2019} together with the fission fragment distributions from \citet[][]{Panov_2001} as our default set of nuclear physics inputs. The neutron capture rates were calculated with the Hauser-Feshbach code \texttt{TALYS} and the photodissociation rates are obtained from detailed balance. For the charged particle reactions the REACLIB rates are used \citep{Cyburt_2010}. In addition, an alternative set of beta-decay rates (D3C*) from \citet[][]{Marketin_2016}, as well as alternative set of fission yields were explored: beta delayed and neutron induced fission fragments from \cite{Mumpower_2020} together with spontaneous fission fragments from \citet[][]{Kodama_1975}. For comparison, the older FRDM1995 mass model \citep{Moller_1995} was also used together with the rates from REACLIB \citep{Cyburt_2010} for all reactions. Finally, two mass models based on energy density functionals were used as described in \citet[][]{Martin_2016} and references therein, again with consistent neutron capture and photodissociation rates: SkM* \citep[][]{Bartel_1982} and UNEDF1 \citep[][]{Kortelainen_2012}\footnote{The mass models SkP, SLy4, SV-min, and UNEDF0 were also tested. They result in similar ratios as the ones in Fig.~\ref{fig:obsVsCalc} but are left out for visual clarity.}.

One has to be careful when comparing the observed and calculated ratios shown in Fig.~\ref{fig:obsVsCalc}, because some stars may also contain an s-process contribution, which could increase the Pt/Eu and Os/Eu ratios, as indicated by the values from the solar s-process residuals. In addition, the third r-process peak elements Pt and Os are strongly affected by the nuclear physics inputs, as evidenced by the large scatter. The position of the third peak can be shifted towards heavier elements depending on nuclear masses, beta decays, and neutron capture cross sections \citep[][]{Arcones_2011,Eichler_2015,Mumpower_2015,Marketin_2016,Lund_2023}. Furthermore, we use a single representative trajectory per astrophysical model, whereas the mass-weighted sum of multiple tracer particles would solidify our results. Even though the observed stars appear to be multi-enriched (see Sec.~\ref{sec:eutail}), we compare their abundances with calculations of individual nucleosynthesis events. This is justified if the r-process is robust and thus produced with same relative abundances by all scenarios. However, we find for some stars a strong enhancement of Os, Ir, and Pt, corresponding to the third r-process peak. This points to a non-robust or additional r-process. Therefore, the observed abundances could be the outcome of a mixture of different r-process patterns. However, even in such a case it is not possible to combine conditions shown in Fig.~\ref{fig:obsVsCalc} and reproduce the highest ratios (magenta dots). In order to explain those ratios, one would need at least one scenario with a high production of third peak elements compared to Eu or Hf and this is not found in any of our conditions. Therefore, the calculated values should be taken with caution. This can be seen by the poor agreement of most calculations with the solar r residuals (solid grey lines) and Sneden's star (brown dot). However, even if the calculations are not conclusive, some clear trends emerge. One of the models based on MRSN (red in Fig.~\ref{fig:obsVsCalc}) tends to produce ratios that are too low to explain the majority of stars (black dots). However, the SkM* calculation provides a closer match to the measured values and the conditions can not be completely ruled out until the nuclear physics is better constrained.

Most stars of this work (black dots in Fig.~\ref{fig:obsVsCalc}) show comparable Os/Eu ratios to those of the robust r-process patterns, represented by solar r residuals and Sneden's star. However, the newly observed ratios show a significantly larger scatter. All newly observed Pt/Eu ratios are at least 0.5-1.0 dex larger than those of the robust r-process patterns. This can be interpreted as a deviation of the robustness, or might hint to a systematic bias towards larger Pt abundances in this work.

Three stars have a distinct enhancement of the third r-process peak abundances (magenta dots in Fig.~\ref{fig:obsVsCalc}). The low metallicities of those stars (see the Eu-tail in Fig.~\ref{fig:ax_vs_eu}) make s-process contributions unlikely - but even the ratios from the solar s residuals (grey dashed lines in Fig.~\ref{fig:obsVsCalc}) would not be high enough to explain the extreme overabundance of the third peak elements in these Eu-poor stars. None of our current models for r-process sites can explain the observed abundances of the third peak of these stars, even considering nuclear physics uncertainties. This points to a different r-process pattern compared to solar or Sneden's star and raises the questions: Is there an additional r-process occurring only at low metallicities, i.e. at very early times, that produces enhanced abundances of third-peak elements but small Eu abundances? Which astrophysical site could explain such new r-process?

The discovery of these abundances further supports the variability of the r-process in contrast to a unique robust pattern\footnote{The un-robustness of the r-process has been discussed before concerning the first r-process peak, low abundances after the second peak \citep[][and references therein]{Hansen:2014aa} and Ac-boost stars \citep[][]{Eichler_2015,Holmbeck_2019}.}
and demonstrates the contribution of an additional r-process at early times. It is puzzling that no currently studied
astrophysical environment can fully
explain the enhanced Os, Ir, and Pt abundances, even when considering nuclear physics uncertainties. A potential new r-process may occur under different conditions than the one producing solar r-process and Sneden-like patterns. This highlights the need for comparing to different observationally derived patterns than those of r-II stars and the Sun.
The nuclear physics uncertainties are still very large and our work clearly evidences the urgent need to reduce them with experiments and theory. With constrained nuclear physics and additional observational data, the third peak elements Ir, Os, and Pt will allow to directly determine the astrophysical conditions of a potential new r-process.

\section{Final remarks}\label{sec:conclusions}

In this work we have performed a homogeneous chemical analysis of 52 stars with high-resolution, high S/N spectra. We have doubled the sample of stellar abundances targeting the third r-process peak (Os, Ir, Pt). We have also demonstrated the viability of doing a large, homogeneous analysis of these species using only ground-based spectra. As a result of exploring such fairly uncharted territory, our study unveiled some intriguing results:

\begin{itemize}
    \item The abundances of the elements from the third r-process peak are decoupled from those of Eu for stars with A(Eu)~$\lesssim$~$-$1.8~dex. They display a flat trend against A(Eu) before starting to follow Eu. That decoupling is not expected from pairs of elements which are supposed to be co-produced in the same conditions.
    \item In most of the stars under study here, the neighbouring species Os (Z-even) and Ir (Z-odd) significantly break the even/odd pattern seen in stellar abundances. That break has been observed in single stars in previous studies, but here we detect it for a larger, homogeneous sample.
    \item The results discussed in the previous two points open the possibility of some unknown mechanism involving an early r-process. It is unlikely that the s-process has some significant influence in these species (Eu, Os, Ir, Pt), in particular at the metallicity range under analysis here.
    \item Surprisingly, the only element in this study that seems to follow Eu is Hf. The behaviour of Hf against Os, Ir, and Pt may suggest that the more metal-rich end of our sample is already experiencing some s-process enrichment, in agreement with results from Paper~III.
    \item This homogeneous study highlights the need for more complete observationally derived abundance patterns at the lowest metallicities (not solar) as well as the importance of careful treatment of the nucleosynthetic yields and their uncertainties.
\end{itemize}

\section*{Data availability}

An external appendix containing five tables is available on Zenodo at \href{https://zenodo.org/records/14225383}{https://zenodo.org/records/14225383}.

\begin{acknowledgements}
      We thank the anonymous referee for the comments and suggestions. We would like to thank M. Hanke, M. Eichler, and M. Molero for help with fruitful discussions and data reduction. AAP, JK, CJH, LL, RFM, and AA acknowledge the support by the State of Hesse within the Research Cluster ELEMENTS (Project ID 500/10.006). CJH also acknowledges the European Union’s Horizon 2020 research and innovation programme under grant agreement No 101008324 (ChETEC-INFRA). AA was supported by Deutsche Forschungsgemeinschaft (DFG, German Research Foundation) -- Project-ID 279384907 - SFB 1245. MR acknowledges support from the Juan de la Cierva program (FJC2021-046688-I) and the grant PID2021-127495NB-I00, funded by MCIN/AEI/10.13039/501100011033 and by the European Union "NextGenerationEU" as well as “ESF Investing in your future”. Additionally, he acknowledges support from the Astrophysics and High Energy Physics program of the Generalitat Valenciana ASFAE/2022/026 funded by MCIN and the European Union NextGenerationEU (PRTR-C17.I1) as well as support from the Prometeo excellence program grant CIPROM/2022/13 funded by the Generalitat Valenciana. EC and PB acknowledge support from the ERC advanced grant N. 835087 -- SPIAKID. This work has made use of data from the European Space Agency (ESA) mission {\it Gaia} (\url{https://www.cosmos.esa.int/gaia}), processed by the {\it Gaia} Data Processing and Analysis Consortium (DPAC, \url{https://www.cosmos.esa.int/web/gaia/dpac/consortium}). Funding for the DPAC has been provided by national institutions, in particular the institutions participating in the {\it Gaia} Multilateral Agreement. This work has made use of the VALD database, operated at Uppsala University, the Institute of Astronomy RAS in Moscow, and the University of Vienna. This work has made use of the \texttt{Python} libraries \texttt{Numpy} \citep[][]{numpy:2020} and \texttt{Matplotlib} \citep[][]{Matplotlib:2007}.
      
\end{acknowledgements}

\bibliographystyle{aa} 
\bibliography{references} 

\begin{appendix} %

\section{Discussion on modelling of NH lines}\label{sec:modelNH}

As an alternative to the discrepant values of oscillator strength found in the literature, as previously discussed in Sect.~\ref{sec:linelist}, we tried a fully theoretical approach in order to derive self-consistent log~gf values for the NH lines of interest in the CERES project. We performed \textit{ab} \textit{initio} molecular calculations on the potential energy curves for the electronic ground $X$ $^3\Sigma$ and $A$ $^3\Pi$ states of NH over a range of internuclear distances from 0.8 up to 10 $\AA$. All the computations were carried out with the Dyall's all-electron core-valence correlated basis set with highest angular momentum equal to 4 (dyall.acv4z). The (small) relativistic effects were modelled by means of the eXact-2-component (X2C) \citep[][]{Peng_2012} correction to the molecular Hamiltonian. The electronic structure was first modelled with the multi-configuration Dirac-Hartree-Fock method, where six valence electrons were distributed over thirteen active atomic orbitals (N's 2s, 2p, 3s and 3p; H's 1s, 2s, 2p). The wavefunctions thus computed were used as a reference for the subsequent Multi-Reference Configuration-Interaction (MRCI) method \citep[][]{Szalay_2011}. The active space included all the NH electrons and accounted for single, double and triple valence excitations. All the calculations were performed with the help of the DIRAC23 \citep[][]{Bast_2023} quantum-chemistry software. Rovibronic transitions were obtained by solving the nuclear Schr{\"o}dinger equation with the help of the software LEVEL \citep[][]{Le_Roy_2017}.

In Table~\ref{tab:A1}, we reported the equilibrium distances $R_e$ and dissociation energies $D_e$ for the electronic ground X $^3$$\Sigma$$^{-}$ and first excited A $^3$$\Pi$ of NH, alongside the related energies for the first vibrational level $E(v = 0)$. For the electronic ground state, our computed internuclear distance falls between the benchmark values of \citet[][]{Fernando_2018} and \citet[][]{melosso_2019}, while the dissociation energy is slightly overestimated (4.9\%) with respect to the former reference. In contrast, our energy for the first vibrational level lies only 0.3\% below the benchmark value of \citeauthor[][]{Fernando_2018}. The accuracy of our spectroscopic parameters further increases with regards to the first excited state, where the discrepancies between our values for $D_e$ and $E(v = 0)$ and the results by \citeauthor[][]{Fernando_2018} amount to 0.95\% and 0.47\%, respectively.

\FloatBarrier

\begin{table}
\caption{Equilibrium distance ($R_e$), dissociation energy ($D_e$) and energy of the first vibrational level for the electronic ground state X $^3$$\Sigma$$^{-}$ and the first excited state A $^3$$\Pi$ of NH.}
\label{tab:A1}
\begin{tabular}{llll}
\hline
\multicolumn{4}{c}{X $^3$$\Sigma$$^{-}$}                                         \\ \hline
                    & $R_e$  & $D_e$  & E($v$ = 0)  \\ 
                    & [$\AA$]       & [cm$^{-1}$]       & [cm$^{-1}$]            \\ \hline
This work           & 1.03684       & 30465             & 1616.497               \\
Koput\tablefootmark{a}           & 1.03591       & 28997             & 1624.010               \\
Owono et al.\tablefootmark{b}    & 1.03983       & 26939             &                        \\
Fernando et al.\tablefootmark{c} &               & 29030             & 1621.450               \\
Melosso et al.\tablefootmark{d}  & 1.03607       &                   & 1623.141               \\ \hline
\multicolumn{4}{c}{A $^3$$\Pi$}                                                  \\ \hline
This work           & 1.01960       & 18289             & 1598.717               \\
Owono et al.\tablefootmark{b}    & 0.98050       & 16712             &                        \\
Fernando et al.\tablefootmark{c} &               & 18466             & 1591.230               \\
Malicet et al.\tablefootmark{e}  & 0.97950       & 16776             & 1590.763               \\ \hline
\end{tabular}%
\tablefoot{
\tablefoottext{a}{\citet[][]{Koput_2015}}
\tablefoottext{b}{\citet[][]{Owono_Owono_2007}}
\tablefoottext{c}{\citet[][]{Fernando_2018}}
\tablefoottext{d}{\citet[][]{melosso_2019}}
\tablefoottext{e}{\citet[][]{Malicet_1970}}}
\end{table}

\begin{table}
\caption{Energies for transition from a lower rovibronic state ($E_{low}$) to an upper rovibronic state ($E_{up}$), alongside the related transition wavelengths in air and log gf.}
\label{tab:A2}
\begin{tabular}{lllll}
\hline
Transition                & E$_{low}$    & E$_{up}$     & $\lambda$ (air) & log~gf \\
                          & [eV]   & [eV]   & [\AA]     &          \\ \hline
X(0,6,e)  $\to$ A(0,7,e)  & 0.082 & 4.085 & 3096.77 & $-$1.521 \\
X(0,7,e)  $\to$ A(0,8,e)  & 0.110 & 4.082 & 3120.39 & $-$1.520 \\
X(0,11,e) $\to$ A(0,10,e) & 0.260 & 4.008 & 3306.97 & $-$1.529 \\ \hline
X(0,6,e)  $\to$ A(0,7,e)  & 0.085 & 3.803 & 3333.11 & $-$1.561 \\
X(0,7,e)  $\to$ A(0,8,e)  & 0.113 & 3.834 & 3330.91 & $-$1.377 \\
X(0,11,e) $\to$ A(0,10,e) & 0.264 & 3.910 & 3399.80 & $-$1.370 \\ \hline
\end{tabular}%
\tablefoot{Transitions are identified by their spectroscopic electronic, vibrational and rotational levels. The reference values by \citet[][]{Fernando_2018} are reported in the second row of this table.}
\end{table}

We further checked the compliance of our calculations with the experimental accuracy requirements by estimating three rovibronic transitions involving the X $^3$$\Sigma$$^{-}$ and A $^3$$\Pi$ states. Besides the aforementioned electronic states, these transitions encompass the $v = 0$ vibrational level, and the $J = 6, 7, 11$ and $J = 7, 8, 10$ rotational levels embedded within both the electronic states. In Table~\ref{tab:A2} we identified these transitions by means of the energy of the lower and upper rovibronic states, the related photon absorption wavelengths in air, and log gf. Our results were compared with the reference data of \citet[][]{Fernando_2018}. While the discrepancies between the two result sets is overall reasonable (between 2 and 7\% on the photon wavelengths in air), they still cannot comply with the 8 $\AA$ experimental resolution. These discrepancies might be reduced by improving the modelling of electron correlation within the intrinsic limits of the MRCI method, by inclusion of higher-order electron excitations; account for non-Born-Oppenheimer effects, following the work of \citet[][]{melosso_2019} should also improve the accuracy of the results. 

\section{Third peak compared to Ba}
\label{sec:Aosvsba}

\begin{figure}[ht!]
    \includegraphics[width=\columnwidth]{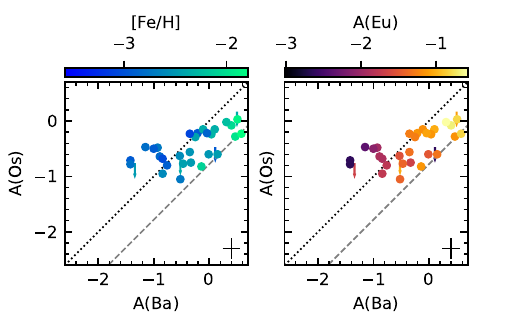}
    \caption{
    A(Os) as function of the A(Ba) abundances calculated in Paper~III for our sample.
    Dotted lines represent the identity function.
    Dashed lines represent solar-scaled values.
    Arrows are upper limits.
    The representative error bars correspond to the median uncertainties.
    }
    \label{fig:osba}
\end{figure}

In Fig.~\ref{fig:osba} we show the behaviour of Os against Ba.

\FloatBarrier
\FloatBarrier

\section{Abundances of Hf, Os, Ir, and Pt}
\label{sec:Aabd}

\begin{table*}[!htbp]
	\centering
	\caption{
 Stellar parameters Teff, log~g, and [Fe/H] from Paper~I, and the abundances calculated in this work in the A(X) scale.}
	\label{tab:abundances}
	\begin{tabular}{lllllrrrr}
		\hline
		              ID                 &     Teff &    log~g &   [Fe/H] & $v_t$ &    A(Hf) &    A(Os) &    A(Ir) & A(Pt) \\
                                       &     [K]  &    [dex] &    [dex] & km s$^{-1}$ & [dex] &   [dex] &    [dex] & [dex] \\
        \hline
  CES0031$-$1647 &     4960 &     1.83 &    $-$2.49 &     1.91 &     $-$1.31 &     $-$0.75 &     $-$0.21 &     0.34 \\
  CES0045$-$0932 &     5023 &     2.29 &    $-$2.95 &     1.76 &     $-$2.08 &     $-$0.71 &     $-$0.56 &     0.39 \\
  CES0048$-$1041 &     4856 &     1.68 &    $-$2.48 &     1.93 &     $-$1.21 &     $-$0.60 &     $-$0.09 &     0.54 \\
  CES0055$-$3345 &     5056 &     2.45 &    $-$2.36 &     1.66 &     $-$1.14 &     $-$0.15 &     $-$0.14 &     0.65 \\
  CES0059$-$4524 &     5129 &     2.72 &    $-$2.39 &     1.56 &    $\cdots$ &     $-$0.35 &        0.08 & $\cdots$ \\
  CES0102$-$6143 &     5083 &     2.37 &    $-$2.86 &     1.75 &    $\cdots$ &    $\cdots$ &    $\cdots$ & $\cdots$ \\
  CES0107$-$6125 &     5286 &     2.97 &    $-$2.59 &     1.54 &    $\cdots$ &    $\cdots$ &        0.32 & $\cdots$ \\
  CES0109$-$0443 &     5206 &     2.74 &    $-$3.23 &     1.69 &     $-$1.30 &    $\cdots$ &     $-$0.15 & $\cdots$ \\
  CES0215$-$2554 &     5077 &     2.00 &    $-$2.73 &     1.91 & $<$ $-$1.77 &     $-$0.48 &     $-$0.36 & $\cdots$ \\
  CES0221$-$2130 &     4908 &     1.84 &    $-$1.99 &     1.84 &     $-$0.74 &     $-$0.28 &     $-$0.26 &     0.65 \\
  CES0242$-$0754 &     4713 &     1.36 &    $-$2.90 &     2.03 &    $\cdots$ &     $-$0.67 &     $-$0.07 & $\cdots$ \\
  CES0301$+$0616 &     5224 &     3.01 &    $-$2.93 &     1.51 &     $-$1.29 &     $-$0.22 &     $-$0.25 & $\cdots$ \\
  CES0338$-$2402 &     5244 &     2.78 &    $-$2.81 &     1.62 &     $-$1.91 &     $-$0.47 &     $-$0.85 &     0.42 \\
  CES0413$+$0636 &     4512 &     1.10 &    $-$2.24 &     2.01 &    $\cdots$ &    $\cdots$ &     $-$0.11 & $\cdots$ \\
  CES0419$-$3651 &     5092 &     2.29 &    $-$2.81 &     1.78 & $<$ $-$1.72 & $<$ $-$0.78 & $<$ $-$0.78 & $\cdots$ \\
  CES0422$-$3715 &     5104 &     2.46 &    $-$2.45 &     1.68 &     $-$1.39 &     $-$0.29 &     $-$0.31 &     0.52 \\
  CES0424$-$1501 &     4646 &     1.74 &    $-$1.79 &     1.74 &        0.11 &     $-$0.23 &     $-$0.28 &     0.82 \\
  CES0430$-$1334 &     5636 &     3.07 &    $-$2.09 &     1.63 &     $-$1.25 & $<$ $-$0.49 &        0.34 &     0.76 \\
  CES0444$-$1228 &     4575 &     1.40 &    $-$2.54 &     1.92 &    $\cdots$ &     $-$0.56 &     $-$0.14 & $\cdots$ \\
  CES0518$-$3817 &     5291 &     3.06 &    $-$2.49 &     1.49 &    $\cdots$ &    $\cdots$ &    $\cdots$ & $\cdots$ \\
  CES0527$-$2052 &     4772 &     1.81 &    $-$2.75 &     1.84 &    $\cdots$ &     $-$1.03 &     $-$0.26 & $\cdots$ \\
  CES0547$-$1739 &     4345 &     0.90 &    $-$2.05 &     2.01 &    $\cdots$ &     $-$0.08 &        0.17 &     1.06 \\
  CES0747$-$0405 &     4111 &     0.54 &    $-$2.25 &     2.08 &    $\cdots$ &     $-$0.02 &    $\cdots$ & $\cdots$ \\
  CES0900$-$6222 &     4329 &     0.94 &    $-$2.11 &     1.98 & $<$ $-$2.14 &     $-$0.82 &     $-$0.22 & $\cdots$ \\
  CES0908$-$6607 &     4489 &     0.90 &    $-$2.62 &     2.12 &    $\cdots$ &     $-$0.95 &    $\cdots$ & $\cdots$ \\
  CES0919$-$6958 &     4430 &     0.70 &    $-$2.46 &     2.17 &    $\cdots$ &     $-$0.77 &    $\cdots$ & $\cdots$ \\
  CES1116$-$7250 &     4106 &     0.48 &    $-$2.74 &     2.14 &    $\cdots$ &     $-$1.05 &    $\cdots$ & $\cdots$ \\
  CES1221$-$0328 &     5145 &     2.76 &    $-$2.96 &     1.60 &    $\cdots$ &    $\cdots$ &    $\cdots$ & $\cdots$ \\
  CES1222$+$1136 &     4832 &     1.72 &    $-$2.91 &     1.93 & $<$ $-$2.43 &     $-$0.80 & $<$ $-$0.28 &     0.17 \\
  CES1226$+$0518 &     5341 &     2.84 &    $-$2.38 &     1.60 &     $-$1.41 &     $-$0.15 &     $-$0.02 &     0.66 \\
  CES1228$+$1220 &     5089 &     2.04 &    $-$2.32 &     1.87 &     $-$0.85 &     $-$0.24 &        0.08 & $\cdots$ \\
  CES1237$+$1922 &     4960 &     1.86 &    $-$3.19 &     1.95 & $<$ $-$1.96 &    $\cdots$ &     $-$0.48 & $\cdots$ \\
  CES1245$-$2425 &     5023 &     2.35 &    $-$2.85 &     1.72 &     $-$1.90 &     $-$0.64 &     $-$0.77 &     0.14 \\
  CES1322$-$1355 &     4960 &     1.81 &    $-$2.93 &     1.96 &     $-$1.57 &     $-$0.50 &     $-$0.47 & $\cdots$ \\
  CES1402$+$0941 &     4682 &     1.35 &    $-$2.79 &     2.01 &    $\cdots$ &     $-$0.78 & $<$ $-$0.41 &     0.33 \\
  CES1405$-$1451 &     4642 &     1.58 &    $-$1.87 &     1.81 &     $-$0.42 &        0.03 &    $\cdots$ &     0.79 \\
  CES1413$-$7609 &     4782 &     1.72 &    $-$2.52 &     1.87 & $<$ $-$2.22 &     $-$0.60 &     $-$0.34 &     0.47 \\
  CES1427$-$2214 &     4913 &     1.99 &    $-$3.05 &     1.85 &     $-$1.98 &     $-$0.68 &     $-$0.46 &     0.25 \\
  CES1436$-$2906 &     5280 &     3.15 &    $-$2.15 &     1.42 & $<$ $-$1.76 &    $\cdots$ & $<$ $-$1.04 &     0.83 \\
  CES1543$+$0201 &     5157 &     2.77 &    $-$2.65 &     1.57 &    $\cdots$ &    $\cdots$ &     $-$0.20 & $\cdots$ \\
  CES1552$+$0517 &     5013 &     2.30 &    $-$2.60 &     1.72 & $<$ $-$2.03 &     $-$0.63 &     $-$0.42 & $\cdots$ \\
  CES1732$+$2344 &     5370 &     2.82 &    $-$2.57 &     1.65 &    $\cdots$ & $<$ $-$0.50 &    $\cdots$ & $\cdots$ \\
  CES1804$+$0346 &     4390 &     0.80 &    $-$2.48 &     2.12 &    $\cdots$ &    $\cdots$ &    $\cdots$ & $\cdots$ \\
  CES1942$-$6103 &     4748 &     1.53 &    $-$3.34 &     2.01 &    $\cdots$ &    $\cdots$ &    $\cdots$ & $\cdots$ \\
  CES2019$-$6130 &     4590 &     1.13 &    $-$2.97 &     2.09 &    $\cdots$ &    $\cdots$ &    $\cdots$ & $\cdots$ \\
  CES2103$-$6505 &     4916 &     2.05 &    $-$3.58 &     1.85 &    $\cdots$ &    $\cdots$ & $<$ $-$0.48 & $\cdots$ \\
  CES2231$-$3238 &     5222 &     2.67 &    $-$2.77 &     1.67 &    $\cdots$ &    $\cdots$ &    $\cdots$ & $\cdots$ \\
  CES2232$-$4138 &     5194 &     2.76 &    $-$2.58 &     1.59 &    $\cdots$ & $<$ $-$0.72 & $<$ $-$1.08 & $\cdots$ \\
  CES2250$-$4057 &     5634 &     2.51 &    $-$2.14 &     1.88 &     $-$1.18 &    $<$ 0.15 &        0.42 & $\cdots$ \\
  CES2254$-$4209 &     4805 &     1.98 &    $-$2.88 &     1.79 &     $-$1.23 &     $-$0.22 &        0.12 & $\cdots$ \\
  CES2330$-$5626 &     5028 &     2.31 &    $-$3.10 &     1.75 & $<$ $-$1.75 &     $-$0.23 &     $-$0.06 & $\cdots$ \\
  CES2334$-$2642 &     4640 &     1.42 &    $-$3.48 &     2.02 &    $\cdots$ &     $-$0.87 &    $\cdots$ & $\cdots$ \\
	    \hline
	\end{tabular}
\end{table*}

\end{appendix}

\end{document}